\documentstyle[12pt]{article}
\newcommand{\beq}{\begin{equation}}
\newcommand{\eeq}{\end{equation}}
\newcommand{\beqa}{\begin{eqnarray}}
\newcommand{\eeqa}{\end{eqnarray}}
\newcommand{\beqar}{\begin{eqnarray*}}
\newcommand{\eeqar}{\end{eqnarray*}}

\renewcommand{\L}{\Lambda}

\renewcommand{\O}{\Omega}

\newcommand{\norm}[1]{\raise.3ex\hbox{:}#1\raise.3ex\hbox{:}}

\def\A{{\cal A}}
\def\B{{\cal B}}
\def\C{{\cal C}}
\def\L{{\cal L}}
\def\M{{\cal M}}
\def\H{{\cal H}}

\def\N{{\cal N}}

\def\O{{\cal O}}
\def\S{{\cal S}}
\def\T{{\cal T}}
\def\G{{\cal G}}
\def\K{{\cal K}}
\def\L{{\cal L}}
\def\R{{\cal R}}

\begin{document}
\begin{titlepage}
\rightline{\small UBC/HEPTH-01}
\vskip 3em

\begin{center}
{\bf \huge
On the Holographic $RG-$flow and the Low-energy,
 Strong Coupling,\\
\vskip 2.5truemm Large $N$ Limit}

\vskip 2em

{\large N. Hambli\footnote{email:
hambli@physics.ubc.ca\hfil}}
\vskip 0.5em

{\em  Department of Physics and Astronomy\\
      6224 Agricultural Road, Vancouver, BC, Canada V6T - 1Z1}

\vskip 3em

\begin{abstract}

From the $AdS/CFT$ correspondence, we learn that the classical evolution of 
supergravity in the bulk can be reduced to a $RG-$flow equation 
for the dual low-energy, strongly coupled and large $N$ gauge theory on the
boundary. This result has been used to obtain interesting relations between
the various terms in the gravitational part of the boundary effective action,
in particular the term that affect the cosmological constant. It is found that
once the cosmological constant is canceled in the $UV$ theory, the $RG-$flow
symmetry of the boundary effective action automatically implies the existence of
zero cosmological constant solutions that extend all the way into the $IR$.
Given the standard (and well founded) contradiction between the $RG-$flow idea
and the observational evidence of a small cosmological constant, this is considered 
to be an important progress, albeit incomplete, towards the final solution. Motivated 
by this success, it would be interesting to see whether this $RG-$stability  
extends outside the scope of strong 't Hooft coupling and large $N$ regime that are implicitly 
assumed in the de Boer-Verlinde-Verlinde Hamilton-Jacobi formulation of the holographic $RG-$flow 
equations of the boundary theory. In this paper, we address this question, where we  
start first by identifying the modifications that are required in the Hamilton-Jacobi formulation of the bulk 
supergravity theory when the strong 't Hooft coupling and the large $N$ limits are relaxed. Next, taking 
into account the leading order corrections in these parameters, we derive new bulk/boundary relations, 
from which one can read all the local terms in the boundary effective action. Finally, we use the resulting 
new constraints, to examine whether the $RG-$stability of the cosmological extends to the new 
coupling regime. It would be also interesting to use these constraints to study the Randall-Sundrum scenario 
in this case. 
\smallskip

\end{abstract}

\end{center}

\end{titlepage}

\setcounter{footnote}{0}
\section{Introduction}

According to the holographic principle [1,2], a macroscopic region of space
and everything inside it can be represented by a boundary theory living on the
boundary of that  region\footnote{Recently, however, the entropy bound on
spacelike and lightlike surfaces has  been generalized to the case of flat
Robertson-Walker geometries in [3] and to more general geometries in [4,5].
See also [6] for work related to the role of focusing mechanism in 
holography.}. Furthermore, the boundary theory should not contain more than
one degree of  freedom per Planck area. This holographic principle has in the
pas few years found a remarkable  realization in superstring theory due to new
insights gained from the investigation of various  superstring dualities. At
the heart of this string theory incarnation of the holographic principle  is
the growing evidence for an intimate connection between quantum phenomena in
gauge theory and  classical aspects of gravity. Early examples illustrating
such a relation are D-branes [7,8], black hole entropy counting [9] and
Matrix theory [10]. However, the clearest statement about  the duality between
gauge theory and gravity is made within the framework of the recently 
discovered ${AdS_D}/{CFT_{D-1}}$ correspondence [11,12,13]. According to this
correspondence the  strong `t Hooft coupling, {\it i.e.}, $g_{YM}^2\, N>>1$,
and the large $N$ limit, {\it i.e.}, $N>>1$,  of certain
$\left(D-1\right)-$dimensional gauge theories have a dual description in terms
of a  supergravity theory defined on one higher-dimensional bulk space. An
important feature of this duality  is the existence of an intriguing
relationship between infrared ($IR$) effects in the bulk theory and 
ultraviolet ($UV$) ones on the boundary. In a succeeding work [14], this
relation was shown to be  crucial in yielding the bound of one degree of
freedom per Planck area as required by the holographic  principle. 
\smallskip 

An immediate follow-up of the $IR/UV$ relation above, which is important to our work in this paper, 
is the interpretation of the extra `radial' $D-$th coordinate $r$, in the bulk space, 
as a renormalization group ($RG$) parameter of the $\left(D-1\right)-$dimensional quantum field theory
living at its boundary. Indeed, the radial evolution of the $D-$dimensional bulk fields was shown to 
share many features with an $RG-$flow [15--19]. This fact was made elegantly
more transparent in the  work of [15] by casting the Einstein equations in the
$D-$dimensional bulk into the form of a Hamiltonian  flow across constant$-r$
{\it timelike} foliations. Specifically, it has been shown that the
Hamilton-Jacobi (HJ) equation for the $D-$dimensional Einstein gravity in the
bulk, with the latter taken to be sliced along  timelike foliations, can be
written in the form of first-order $RG-$flow equations of the classical 
supergravity action. Furthermore, in the {\it asymptotic} limit where the $UV$
boundary extends all the  way to infinity, these $RG-$flow equations reduce to
the standard Callan-Symanzik equation including the  the conformal anomaly
terms [20], in full accordance with the $RG-$ flow ideas in quantum field
theory. This result lends support to the identification of the bulk classical
supergravity action with the  boundary quantum effective action of the gauge
theory as suggested in [11,13].   
\smallskip

In the standard ${AdS_D}/{CFT_{D-1}}$ correspondence (as described above), where the bulk spacetime 
is taken to be non-compact, the dual boundary theory is at infinite bulk radius. As such, it must have 
an infinite energy $UV$ cutoff by virtue of the $IR/UV$ relation. Therefore, the $D-$dimensional bulk 
graviton modes that extend all the way to the $UV$ boundary are {\it not normalizable}, and hence 
gravity decouples totally form the boundary, leaving out there pure Yang-Mills theory. However, as first 
pointed out in [21] by Randall and Sundrum (RS), this situation changes as
soon as one considers the transverse  bulk radius to be of {\it finite} range.
This in effect translates into having a dual boundary theory at finite  bulk
radius, and hence with a {\it finite} $UV$ cutoff due to the $IR/UV$ relation.
In this case, there  will exist {\it normalizable} fluctuations of the
$D-$dimensional metric that propagate and couple as  graviton modes of the
$\left(D-1\right)-$dimensional boundary theory. This generalization of  the
${AdS_D}/{CFT_{D-1}}$ correspondence leads also to a remarkable interplay
between Einstein  equations of the coupled gravity-matter theory on the
boundary and the $RG-$flow equations [15,16,17,21]. In addition, it provides
interesting relations between the various terms in the  boundary quantum
effective action, in particular the boundary Newton constant, the cosmological
constant and the scalar potential [15,16,17,21]. As a result, a cosmological
constant is naturally  prevented from being generated dynamically along the
$RG-$flow once it has been canceled at higher  energies inside the bulk, as
pointed out in [15]. These results join and corroborate earlier findings  on
the role of large extra dimensions in the resolution of the cosmological
constant puzzle [22]. 
\smallskip

In principle, the above results should continue to hold for any $\left(D-1\right)$-dimensional 
gauge theory provided that it can be represented as a relevant or marginal perturbation 
(in the sense of [23]) of a large $N$ superconformal field theory or any
deformation of it, for  which the ${AdS_D}/{CFT_{D-1}}$ correspondence has
been established. It is important to point out,  though, that two main
assumptions went into the derivation of the $RG-$flow equation of the boundary
 gauge theory from the HJ equation of the classical supergravity action in the
bulk, as presented first in [15]. These two assumptions are, inherently, part
of the conditions that are involved in the derivation ${AdS_D}/{CFT_{D-1}}$
correspondence. The first assumption concerns  the requirement that the gauge
theory must have a large $N>>1$, (and thus a large gauge group)  so that one
can neglect the string loop effects represented by the $1/{N^2}$ corrections.
Secondly, the gauge theory is required to have a large `t Hooft coupling,
$g_{YM}^2\, N>>1$, which amounts to taking the energy scale in the theory to
be low enough so that one can ignore  quantum gravity effects controlled by
the `stringy' ${\alpha'}/{R^2}$ corrections. ${\alpha'}$  denotes as usual the
square of the string length, and $R$ represents some characteristic radius of 
the bulk geometry\footnote{To understand better these limits, we refer the
reader to section~(2)  where we show that for type $IIB$ superstrings on
$AdS_5 \times S^5$, the string coupling  is $g_{st} \sim g_{YM}^2 \sim
1/{N^2}$, and the radius is $R^2 \sim \alpha' \, \sqrt{g_{YM}^2\, N}$.}.
Therefore, one expect to have significant modifications of the HJ equation and
 hence the $RG-$flow equations outside this low-energy, strongly coupled,
large $N$ regime.  It is the purpose of this paper to identify the changes
that are brought in the derivation  of the $RG-$flow equation from the
equation when the limits $N>>1$ and ${\alpha'}/{R^2}<<1$ are  relaxed. In
other words, we are interested in the calculation of the leading-order
corrections, in the  parameters $1/N$ and ${\alpha'}/{R^2}$, to the HJ
equation of the bulk supergravity, and in the study  of their consequences. 
\smallskip

We start in section~(2) by reviewing briefly the ${AdS_D}/{CFT_{D-1}}$ correspondence to set 
notation and especially to emphasis the emergence of the large $N$ and large `t Hooft coupling.
In section~(3), we introduce the leading $\alpha'$ corrections in the bulk supergravity action 
[24,25,26]. These corrections have their origin in the vanishing of the beta
function of the string  theory non-linear sigma model. They are represented by
higher-derivative local effective interactions  involving the higher-curvature
gravitational terms. Next, we give a Hamiltonian formulation  of the the bulk
higher-curvature supergravity action so obtained. As expected, we find  that
the HJ equations are changed since the canonical conjugate momentum to the 
metric inherits in this case new terms coming from the $\alpha'$ corrections.
Even though, it is  tedious to calculate the changes that are brought by the
$\alpha'$ corrections to the  HJ equations, their form and how they appear as
higher-derivative non-renormalizable  effective interactions can be derived
systematically in string theory using effective field theory  language
[23,27]. In section~(4), we deal with the question of how to incorporate the
$1/N$  corrections in the HJ equations. There is a striking similarity between
our problem here and the one  we face when we make the transition from the
{\it classical} HJ equations to the {\it quantum} Schr{\"{o}}dinger equation. In
that context, using the $WKB$ or semiclassical theory, the leading quantum
corrections linear  in $\hbar$ are found to be proportional to the second
order variation of the action $S$. In a similar manner,  the $1/N$ corrections
which would change the $RG-$flow equations are taken to be represented by
second order  variations of the supergravity bulk action. The interpretation
of the HJ constraints of the bulk theory as  giving us the $RG-$flow equations
of the boundary theory taken at the radius where the HJ constraints are 
satisfied, rests also upon their strong resemblance with Polchinski's exact
$RG$ equation [28]. Therefore,  in section~(4), we also use this connection to
motivate the addition of the second order variations of the action as
representing the $1/N$ corrections. After adding the $\alpha'$ and $1/N$
corrections,  we look in section~(5) for their implication on the relations
between quantities in the boundary action  previously derived in [15,17], in
particular those involving the Newton constant, the scalar potential  and the
cosmological constant. Furthermore, it would be interesting to see whether the
solution to the  cosmological constant problem as proposed in [15,17] is
affected in this case. Finally, in section~(7), we discuss our results and offer
suggestions for future directions. The Hamiltonian formulation of general
relativity in  the presence of higher-curvature terms is presented in the
appendix. 
\smallskip   

\section{${AdS_D}/{CFT_{D-1}}$ correspondence and holography }

We start by reviewing quickly some basic elements of the $AdS_D/CFT_{D-1}$ correspondence. 
Our main concern here will be to motivate the large $N>>1$, and the large 't Hooft coupling
$g_{YM}^2\, N >> 1$ limit, involved in the correspondence. Furthermore, to simplify our presentation, 
we focus only on the $D=5$ case since many of the features found in this case continue to 
hold for general $D$. The most studied example in this category is the proposed duality between $4-$dimensional 
Yang-Mills theory with $\N = 4$ supersymmetries and type $IIB$ superstring theory on $AdS_5 \times S^5$ 
geometry. At the heart of this duality is the existence of the relation between the two different descriptions 
of a stack of $N$ parallel extremal D3-branes. One in terms of the low-energy $\left(4\right)-$dimensional 
$U\left(N\right)$, $\N = 4$ supersymmetric gauge theory on its world-volume, and the other in terms of the 
classical supergravity background of the type $II$ closed superstring theory. An essential step 
in the derivation of the $AdS_5/CFT_4$ correspondence is the understanding of the range of validity of
each of the description above. For the classical supergravity description, we need the form of the background string
metric, the dilaton and the $RR-$gauge field for the stack of $N$ parallel extremal D3-branes. 
This is given by the following form
\beqa
ds^2 & = & \left(1 + {R^4}/{r^4}\right)^{-\, {1\over 2}}\, d x_{//}^2 +
\left(1 + {R^4}/{r^4}\right)^{1\over 2}\, \left(d r^2 + r^2 \, d \Omega_5^2 \right)\, ,
\label{ba}\\
e^{\phi} & = & g_{st}\, ,
\label{bb}\\
C_{0123} & = &  \left(1 + {R^4}/{r^4}\right)^{-\, 1} - 1\, ,
\label{bc}
\eeqa
where $d x_{//}^2$ denotes the flat $4-$dimensional metric for the coordinates parallel to the D3-branes,
and the radius $R$ is $R^2 = \alpha' \, \sqrt{g_{st}\, N}$. For the low-energy supersymmetric Yang-Mills 
description on the D3-branes world-volume, we need the relation $g_{YM}^2 = g_{st}$ between the 
couplings\footnote{For a general Dp-brane, the relation between the couplings is 
$g_{YM}^2 = g_{st}\, \left(\alpha'\right)^{p-3}$, and the dimensionless effective coupling, at energy scale 
$E$, is $g_{\hbox{eff}}^2 \left(E\right) = g_{YM}^2\, N\, E^{p-3}$. Perturbation theory applies in 
the $UV$ for $p<3$, and in the $IR$ for $p>3$, and the two cases may be
related by S-duality [12]. The special case $p=3$, presented in section~(2),
corresponds to $\N =4$ supersymmetric Yang-Mills  theory in $D=4$, which is
known to be a finite, conformally invariant quantum field theory.}. 
\smallskip

Another piece of knowledge which played an important role in the formulation of the standard $AdS_5/CFT_4$ 
correspondence is the realization that the low-energy limit of the gauge theory on the D3-branes world-volume, 
corresponding to $\alpha' \to 0$, may be taken directly in the supergravity description. On the supergravity 
side, the limit amounts simply to taking the near horizon geometry corresponding to the $r \to 0$ limit. Thus, 
finally, in the limit $\alpha' \to 0$ and $r \to 0$, with $r/\alpha'$ fixed, one finds that the metric in 
(\ref{ba}) reduces to the form 
\beq
ds^2 = {{r^2}\over{R^2}} \, d x_{//}^2 + {{R^2}\over{r^2}} \, d r^2 + R^2 \, d \Omega_5^2 \, ,
\label{bd}
\eeq 
which describes the product-space geometry $AdS_5 \times S^5$, where both factors have radius 
$R^2 = \alpha' \, \sqrt{g_{YM}^2 \, N}$. Furthermore, we know that the classical supergravity 
description can be trusted only if the length scale of the D3-brane solution, given by the metric 
(\ref{bd}), is much lager than the string scale $\sqrt{\alpha'}$, which allows for the `stringy' 
quantum gravity effects to be neglected. This condition translates into $R^2 >> \alpha'$, 
which yields the large 't Hooft coupling limit for the gauge theory on the D3-branes world-volume, 
{\it i.e.}, $g_{YM}^2\, N >> 1$. In order to suppress the string loop corrections, we also need to
take $g_{st} \to 0$, and hence $g_{YM}^2 \to 0$, which amounts to taking the large $N$ limit, $N>>1$. 
To summarize, the supergravity solution is expected to give exact information about the $\N = 4$ 
supersymmetric Yang-Mills theory on the D3-branes world-volume, in the limit of large $N>>1$ and 
large `t Hooft coupling $g_{YM}^2\, N >>1$. More on the two limits above after introducing another
key feature of the $AdS_5/CFT_4$ correspondence below, that is, the idea of $RG-$ flow and holography. 
\smallskip

From the $AdS_5 \times S^5$ geometry in (\ref{bd}), we can see that the coordinate $r$ transverse 
to the D3-branes can be regarded as a renormalization group scale. Indeed, two excitations in the
gauge theory on the D3-branes world-volume, which are related by a scale transformations
\beq
x_{//} \to e^{\tau}\, x_{//}\, ,
\label{be}
\eeq
translate on the $AdS-$factor of the geometry into two excitations concentrated around
different locations in the transverse $r-$direction, and which are related by the following 
transformation [12,29]
\beq
r \to e^{-\, \tau}\, r \, .
\label{bf}
\eeq
The $AdS_5/CFT_4$ correspondence provides us thus with a holographic map between physics in the gauge
theory on the world-volume, which can be thought of as living on the ${AdS_5}-$boundary,
and physics in one higher dimension in $AdS_5$ bulk space. This holographic map is at the center
of the $IR/UV$ relation according to which ($IR$) effects in the bulk theory are related
to ($UV$) ones on the boundary. This relation turned out to be very crucial in yielding the 
holographic bound of one degree of freedom per Planck area as required by the holographic principle 
[15]. 
\smallskip

In the original $AdS_5/CFT_4$ correspondence, the $AdS_5$ boundary is taken to be at $r = +\, \infty$, 
and as a result the range of the $r-$values extends all the way to infinity. Therefore, while the theory 
in the $AdS_5$ bulk space contains gravity, the dual $CFT_4$ theory on the boundary does not. This happens 
because the bulk gravitational modes that propagate all the way to infinity are not normalizable, and 
therefore do not fluctuate. In this paper, however, we are interested in the much more general 
situation where gravity does not decouple at the boundary. For this to happen, we follow the Randall-Sundrum 
proposal in [21], and choose the $AdS_5$ transverse $r-$coordinate to run
over a finite range, $r\le r_0$, instead over an infinite range. An immediate
consequence of this is that, there exists now a normalizable gravitational 
collective mode at the boundary, which in this case is living at finite the
radius $r = r_o$. Furthermore, in view of the $IR/UV$ relation,  truncating
the bulk theory to $r-$values larger (or smaller) than some finite $r = r_o$
amounts to introducing  a finite $UV$ (or $IR$) cut-off in the theory at the
boundary [15]. Therefore, allowing for the bulk  transverse $r-$direction to
be interpreted as an $RG$ scale. Indeed, by casting the bulk  Einstein
equations into the form of Hamiltonian flow across timelike boundaries, the 
$r-$evolution of the bulk fields were shown in [15] to share many features
with an $RG-$ flow  on the boundary.  
\smallskip

Combining this holographic perspective of $AdS_5/CFT_4$ correspondence with the $RG$ scale interpretation 
of the bulk transverse $r-$coordinate, one aims to derive the low-energy quantum 
effective action, $\S_{\hbox{b}}$, on the boundary from the knowledge of the bulk  
supergravity theory. As explained above, we shall take the boundary 
to be at finite radius $r_o$ so that gravity does not decouple from the boundary theory. 
To this end, we start by defining some classical action for the supergravity theory in the bulk, 
which we denote by $S_T\, \left[\phi^I , g\right]$. Besides the bulk metric,
$\G_{AB}$, $S_T\, \left[\phi^I , g\right]$ depends also on some scalar fields $\phi^I$ that represent 
the various couplings of the boundary theory. In fact, it is the evolution of these 
scalar fields as a function of the bulk transverse $r-$coordinate that eventually lead to the  
$RG-$flow equations on the boundary theory\footnote{Due to the stress energy-momentum tensor of the scalar fields 
$\phi^I$, the background geometry in the bulk will deviate from that of a pure $AdS_5$ form.}. 
For later reference, we choose the bulk metric to be of the form
\beq
ds^2 = \G_{AB}\, dx^A\, dx^B = \left( N^2 + N_\mu\, N^\mu \right)\, dr^2 + 2\, N_\mu\, d x^\mu\, dr
+ g_{\mu\nu}\, \left(x,r\right)\, dx^{\mu}\, dx^{\nu}\, ,
\label{bg}
\eeq
where the upper case Latin letters, $A$ and $B$, are taken to denote the bulk coordinates 
$\left(r , x^\mu\right)$, and the lower case Greek indices, $\mu$ and $\nu$ denote 
the boundary coordinates. We assume the boundary metric $g_{\mu\nu}\, \left(x,r\right)$ to be
of Euclidean signature, and we allow the scalars $\phi^I\, \left(x,r\right)$ to depend on all bulk 
coordinates $\left(r, x^\mu\right)$. $N$ and $N^\mu$ are the lapse and shift functions, respectively. 
A convenient choice of coordinates are the Gaussian normal coordinates, where $N^\mu = 0$ and $N=-1$. 
Using such coordinates, the metric in (\ref{bg}) takes on the simple form
\beq
ds^2 = dr^2 + g_{\mu\nu}\, \left(x,r\right)\, dx^{\mu}\, dx^{\nu}\, .
\label{bh}
\eeq
(More details on our notation and convention are presented in the appendix section).
\smallskip

Finally, one of the main ingredients in the $AdS_5/CFT_4$ correspondence is the identification of the 
classical supergravity action $S_T\, \left[\phi^I , g\right]$ evaluated on a classical solution, with 
specified boundary values $g_{\mu\nu}\, \left(x, r_o\right)$ and $\phi^I\, \left(x, r_o\right)$, with
the generating functional of gauge invariant correlators of gauge invariant observables $\O_{I}$ in 
the boundary theory living at $r=r_o$. that is, we have 
\beq
\left .
\left\langle\, \O_{I_1}\, \left(x_1\right)\, ...\, \O_{I_n}\, \left(x_n\right)\, \right\rangle =
{1\over\sqrt{g\left(x_1\right)}}\, {\delta\over{\delta\, \phi^{I_1}\, \left(x_1\right)}}
\, ...\,
{1\over\sqrt{g\left(x_n\right)}}\, {\delta\over{\delta\, \phi^{I_n}\, \left(x_n\right)}}\,
S_T\, \left[\phi^I , g\right]\right|_{r=r_o}\, .
\label{bi}
\eeq
By requiring that the scalar fields $\phi^I$ and the metric $g_{\mu\nu}$ stay regular
inside the bulk, there is in principle one unique supergravity classical solution
for a given boundary value for $\phi^I$ and $g_{\mu\nu}$.
If we put the scalar fields $\phi^I$ to zero after doing the variation, we do
obtain the gauge invariant correlators of the unperturbed $\N =4$ supersymmetric Yang-Mills 
boundary theory. If the fields
$\phi^I$ are put to finite values, however, the resulting 
boundary theory will correspond to a finitely perturbed $\N =4$ supersymmetric
Yang-Mills theory.
\smallskip

Although the discussion, in this section, was so far limited to the $AdS_5/CFT_4$ correspondence, one 
could easily generalize it to include the higher-dimensional $AdS_D$ spaces. We would be then talking  
about an $AdS_D/CFT_{D-1}$ correspondence. In similarity with the $AdS_5/CFT_4$ correspondence, 
the large $N$ limit, $N>>1$, and the large `t Hooft coupling limit, ${R^2}/\alpha' = \sqrt{g_{YM}^2 \, N} >>1$ 
will also be involved in this case. In particular, the interpretation of the radial $AdS_D$ coordinate
with an $RG$ scale will also allow in this case for an identification of the radial evolution of the bulk 
fields with a $RG-$flow. Thus, by working within the general framework of $AdS_D/CFT_{D-1}$
correspondence, our purpose next will be to go beyond the large $N$, 
and large ${R^2}/\alpha'$ limit, and consider the leading corrections in $1/N$ and $\alpha'/{R^2}$ 
to the $RG-$flow equations derived from the bulk HJ constraint. 
\smallskip

\section{HJ equations and the higher-curvature terms}

In this section, we consider the derivation of the HJ constraint of the $D-$dimensional bulk 
supergravity theory in the presence of the $\alpha'$ corrections coming from a quantum theory 
of gravity such as string theory. In string theory, the lowest-order $\alpha'-$corrections to the low-energy 
effective action involve the higher-curvature terms, which are controlled by the expansion parameter 
$\alpha'/{R^2}$, where $R$ is the characteristic radius of the bulk space. Therefore, by virtue of the relation 
$R^2/\alpha' = \sqrt{g_{YM}^2\, N}$, the addition of the higher-curvature terms will necessarily affect the 
large 't Hooft coupling limit $g_{YM}^2\, N >>1$ involved in the $AdS_D/CFT_{D-1}$ correspondence as well the
the $RG-$ flow equations derived from it. As in the approach of [15,17], the
changes that are brought by  the higher-curvature corrections are most
conveniently analyzed using the HJ theory of the $D-$dimensional  bulk
supergravity theory. By casting the Einstein's equations in the bulk  into the
form of a Hamiltonian evolution across timelike boundaries,  one is led to the
familiar HJ constraint of the canonical formalism of  gravity. It has been
shown in [15,17] that this constraint play a key role in the bulk/boundary 
correspondence, as they allow for a systematic derivation of the
$\left(D-1\right)-$dimensional  quantum effective action of the boundary
theory from the knowledge of the bulk theory. Furthermore,   combining these
results with the $IR/UV$ relation, a holographic $RG-$flow picture of the dual
boundary  theory naturally emerges, as changes in the bulk coordinate $r=r_o$
translates now  into shifts in the energy scale $\mu = r/\alpha'$ of the dual
boundary theory. 
\smallskip

For the purpose of deriving the changes to the HJ constraint due to the higher-curvature corrections,
we choose for the $D-$dimensional bulk spacetime the following supergravity action\footnote{Since we 
choose to work within the Einstein frame, we can use the $D-$dimensional Planck unit such that $\kappa_D = 1$.}
\beq
S_T = S_1 + S_2 + S_3\,,
\label{ca}
\eeq
where $S_1$, $S_2$ and $S_3$ are given by 
\beqa
S_1 & = & \int_D\, \sqrt{\G}\, dr\, {d^{D-1}}x\, \left[{}^{\{D\}}R + 2\, \Lambda + a_o\, {}^{\{D\}}R^2 + 
b_o\, {}^{\{D\}}R_{\mu\nu}\, {}^{\{D\}}R^{\mu\nu} \right.
\nonumber\\
& & 
\left. \qquad\qquad\qquad\qquad\qquad\qquad\qquad\qquad\qquad 
+ c_o\, {}^{\{D\}}R_{\mu\nu\rho\sigma}\, {}^{\{D\}}R^{\mu\nu\rho\sigma}\right]\, ,
\label{cba}\\
S_2 & = & 2\, \int_{D-1}\, \sqrt{g}\, {d^{D-1}}x\, \left[\, K + {{\K}}\left(K, \nabla\, K ; a,b,c\right)\, 
\right]\, ,
\label{cbb}\\
S_3 & = & \int_{D}\, \sqrt{\G}\, dr\, {d^{D-1}}x\, \left[\, V\left(\phi^I \right) - 
{1\over 2}\, \nabla_\mu\, \phi^I\, \nabla^\mu\, \phi^I\, \right]\, .
\label{cbc}
\eeqa
Besides the curvature action $S_1$, the $D-$dimensional bulk supergravity theory contains 
the matter action $S_3$ for the bulk scalar fields $\phi^I$, which through the bulk/boundary
correspondence represent the various coupling of the dual boundary theory. The sum over repeated
index $I$ of the bulk scalar fields is assumed throughout all the paper. The curvature scalar 
${}^{\{D\}}R$, and the tensors ${}^{\{D\}}R_{\mu\nu}$ and ${}^{\{D\}}R_{\mu\nu\rho\sigma}$ appearing 
in the action $S_1$ are calculated using the $D-$dimensional bulk metric $\G_{AB}$, with the 
coefficients $a_o$, $b_o$ and $c_o$ parameterizing for the time being only the $\alpha'/{R^2}$ 
corrections\footnote{Later, in section~(4), the coefficients of the higher-curvature terms will 
include, besides $a_o$, $b_o$ and $c_o$, the contributions $a_1$, $b_1$ and $c_1$ from the 
$1/N$ corrections.}. Using the Gaussian normal coordinates, $\G_{AB}$ takes the form
\beq
ds^2 = \G_{AB}\, dx^A\, dx^B = dr^2 + g_{\mu\nu}\, \left(x,r\right)\, dx^{\mu}\, dx^{\nu}\, .
\label{cc}
\eeq
Following section~(1), our notation will be to take the upper case Latin letters such as $A$ and $B$ to 
denote the $D-$dimensional bulk coordinates $\left(x^\mu ,r\right)$, where the lower case Greek indices such 
as $\mu$ and $\nu$ are taken to denote the coordinates $\left(x^\mu ; \mu = 0,1,...,D-2\right)$ of the 
$\left(D-1\right)-$dimensional boundary theory. As usual, $K$ is taken to denote the extrinsic curvature of the 
boundary surface, whose form is given by 
$K_{\mu\nu} = {1\over 2}\, {\L_r}\, g_{\mu\nu} = {g_\mu}{}^\rho \, \nabla_\rho\, n_\nu$.
Whereas ${{\K}}\left(K, \nabla\, K ; a,b,c\right)$ is taken to represent the additional surface terms 
corresponding to the higher-curvature terms in $S_1$. The derivation of ${{\K}}\left(K, \nabla\, K ; a,b,c\right)$ 
was carried out in [24,25], but its exact form will not be necessary for our
work here. It suffices to recall   that the normal-derivative terms from $S_1$
are canceled by the variation of the surface terms in $S_2$. 
\smallskip

Using the choice of the metric in (\ref{cc}), the $D-$dimensional bulk equations of motion 
can be cast into a form of a Hamiltonian flow across the $\left(D-1\right)-$dimensional
timelike boundaries, with the bulk transverse $r-$coordinate playing the role of time.
Based on the bulk action $S_T$ in (\ref{ca}), the Hamiltonian that generates this radial flow 
is explicitly derived in the appendix section, where we found  
\beqa
-\,\, {1\over\sqrt{g}}\,\, \H_T & = & \left( \R + 2\, \Lambda + a_o\, \R^2 + b_o\, \R_{\mu\nu}\, \R^{\mu\nu} 
+\, c_o\, \R_{\mu\nu\rho\sigma}\, \R^{\mu\nu\rho\sigma} \right)
\nonumber\\
& & 
\, + \, \left( V \left(\phi\right)\, -\, 
{1\over 2}\, \nabla_\mu\, \phi^I\, \nabla^\mu\, \phi^I \right)
\nonumber\\
& & 
\, +\, \left( \pi_{\mu\nu}\, \pi^{\mu\nu} - {{\pi^2}\over{D - 2}} \right)
\, +\, {1\over 2}\, \pi_I\, \pi_I
\nonumber\\  
& &
\, -\, 2\, a_o\, \R\, \left( \pi_{\mu\nu}\, \pi^{\mu\nu} - {{\pi^2}\over{D - 2}} \right)
\, - \, 2\, b_o\, \R_{\mu\nu} \, \left( {\pi^{\mu}}_\rho\, \pi^{\rho\nu} - 
{{\pi\, \pi^{\mu\nu}}\over{D - 2}} \right)
\nonumber\\
& &
\, + \, 4\,\, c_o\, \R_{\mu\nu\rho\sigma}\, \pi^{\mu\rho}\, \pi^{\nu\sigma}
\, + \, {{4\, c_o}\over{\left(D - 2\right)^2}}\, \R\, \pi^2
\, -\, {{8\, c_o}\over{D - 2}}\, \R_{\mu\nu}\, \pi\, \pi^{\mu\nu}\, ,
\label{cd}
\eeqa
with $\pi_{\mu\nu}$ and $\pi_I$ are the canonical momentum variables conjugate to $g^{\mu\nu}$ and $\pi^I$, 
respectively. It is a standard fact, well known in classical mechanics, that 
(given the bulk action $S_T$) the value of the canonical momentum $\pi_I$ conjugate to $\phi^I$, at a 
given slice $r = \hbox{const}$, is equal to the functional derivative of the bulk action 
$\S_{\hbox{b}}$ induced on that slice, with respect to $\phi^I$. Furthermore, since the conjugate momentum 
$\pi_I$ is related to the radial flow of $\phi^I$, we have
\beq
{1\over\sqrt{g}}\,  {{\delta\, \S_{\hbox{b}}}\over{\delta\, {{\phi^I}}}} 
= \pi^I = \L_r\, \phi^I = -\, \dot{\phi^I}\, .
\label{cg}
\eeq
Similarly, for the canonical momentum variable $\pi^{\mu\nu}$ conjugate to the the metric $g_{\mu\nu}$,
we have 
\beqa
{1\over\sqrt{g}}\, {{\delta\, \S_{\hbox{b}}}\over{\delta\, {{g}}^{\mu\nu}}} = \pi_{\mu\nu}  
& = & \left( K_{\mu\nu} - K\, g_{\mu\nu} \right) + 
2\, a_o\, \R \, \left( K_{\mu\nu} - K\, g_{\mu\nu} \right) 
-\, b_o\, \left( K\, \R_{\mu\nu} \, +\,  \R^{\rho\sigma}\, K_{\rho\sigma}\, g_{\mu\nu} \right) 
\nonumber\\
& & 
+\, b_o\, \left( {\R_\mu}^\rho \, K_{\rho\nu} \, +\, {\R_\nu}^\rho \, K_{\rho\mu} \right)
-\, 4\, c_o\, \R_{\mu\rho\nu\sigma}\, K^{\rho\sigma}\, +\, \O \left( K^3 \right)\, ,
\label{ch}
\eeqa
or by taking the trace
\beqa
\pi = g^{\mu\nu}\, \pi_{\mu\nu} = {\pi^\mu}_\mu & = & -\, \left( D - 2 \right)\, K
\, - \, \left( 2\, a_o\, D - 4\, a_o + b_o \right)\, \R\, K
\nonumber\\
& & 
\, - \, \left( b_o\, D - 3\,  b_o + 4\, c_o \right)\, \R_{\mu\nu}\, K^{\mu\nu}\, 
+\, \O \left( K^3 \right)\,  .
\label{ci}
\eeqa
(The details about the calculation of the conjugate momenta $\pi_I$ and $\pi^{\mu\nu}$ can be
found in the appendix section.) 
\smallskip

Hamiltonian flow across the boundary is a constrained system, since it is still endowed with 
redundancies. The choice of the foliation is arbitrary, and even after fixing one, the system 
is still endowed with redundancies. To remove completely these redundancies, two set of constraint 
equations on the initial data at the boundary are necessary. The first constraint translates simply 
into a statement regarding Poincar\'e invariance on the boundary slices. It ensures that the boundary 
effective action is invariant under $\left(D-1\right)-$dimensional coordinate transformations. 
The second constraint, which is most important for our purposes, is the Hamilton constraint. It
requires  to set $\H_T = 0$, which ensures invariance of the constant$-r$ slices under the local shifts.
Using (\ref{cd}) for the Hamiltonian $\H_T$, we obtain 
\beqa
\left( {{\pi^2}\over{D - 2}} - \pi_{\mu\nu}\, \pi^{\mu\nu} \right)
\, -\, {1\over 2}\, \pi_I\, \pi_I
\, +\, 2\, a_o\, \R\, \left( \pi_{\mu\nu}\, \pi^{\mu\nu} - {{\pi^2}\over{D - 2}} \right) 
\, + \, 2\, b_o\, \R_{\mu\nu} \, \left( {\pi^{\mu}}_\rho\, \pi^{\rho\nu} - 
{{\pi\, \pi^{\mu\nu}}\over{D - 2}} \right) &&
\nonumber\\
 - \, 4\,\, c_o\, \R_{\mu\nu\rho\sigma}\, \pi^{\mu\rho}\, \pi^{\nu\sigma}
\,- \, {{4\, c_o}\over{\left(D - 2\right)^2}}\, \R\, \pi^2
\, +\, {{8\, c_o}\over{D - 2}}\, \R_{\mu\nu}\, \pi\, \pi^{\mu\nu} = &&
\nonumber\\
\left( \R + 2\, \Lambda + a_o\, \R^2 + b_o\, \R_{\mu\nu}\, \R^{\mu\nu} 
+\, c_o\, \R_{\mu\nu\rho\sigma}\, \R^{\mu\nu\rho\sigma} \right)
\, + \, \left( V \left(\phi\right)\, -\, 
{1\over 2}\, \nabla_\mu\, \phi^I\, \nabla^\mu\, \phi^I \right)\, .
\label{cf}
\eeqa
\smallskip

To obtain the HJ constraint at $r = r_o$, we simply have to replace the canonical momenta in (\ref{cf}) 
by the functional derivatives of the bulk action ${\S_{\hbox{b}}}$ induced on $r = r_o$, with respect to the 
conjugate variables. In terms of the action ${\S_{\hbox{b}}}$, the HJ constraint reads  
\beqa
 2\, a_o\, {\R\over\sqrt{g}}\, 
\left[{{\delta\, {\S_{\hbox{b}}}}\over{\delta\, g_{\mu\nu}}}\, 
{{\delta\, {\S_{\hbox{b}}}}\over{\delta\, g^{\mu\nu}}}
-\, {1\over{D-2}}\, \left(g_{\mu\nu}\, 
{{\delta\, {\S_{\hbox{b}}}}\over{\delta\, g_{\mu\nu}}}\right)^2\right] 
+\, 2\, b_o\, {\R_{\mu\nu}\over\sqrt{g}}\,
\left[{{\delta\, {\S_{\hbox{b}}}}\over{\delta\, {{g_\mu}^\rho}}}\, 
{{\delta\, {\S_{\hbox{b}}}}\over{\delta\, g_{\rho\nu}}}
-\, {{g_{\rho\sigma}}\over{D-2}}\, {{\delta\, {\S_{\hbox{b}}}}\over{\delta\, g_{\rho\sigma}}}\,
{{\delta\, {\S_{\hbox{b}}}}\over{\delta\, g_{\mu\nu}}}\right] & &
\nonumber\\
-\, 4\, c_o\, {\R_{\mu\nu\rho\sigma}\over\sqrt{g}}\, 
{{\delta\, {\S_{\hbox{b}}}}\over{\delta\, g_{\mu\rho}}}\,
{{\delta\, {\S_{\hbox{b}}}}\over{\delta\, g_{\nu\sigma}}}
-\, {{4\, c_o}\over{\left(D-2\right)^2}}\, {\R\over\sqrt{g}}\,
\left(g_{\mu\nu}\, {{\delta\, {\S_{\hbox{b}}}}\over{\delta\, g_{\mu\nu}}}\right)^2
+\, {{8\, c_o}\over{D-2}}\, {\R_{\mu\nu}\over\sqrt{g}}\,
g_{\rho\sigma}\, {{\delta\, {\S_{\hbox{b}}}}\over{\delta\, g_{\rho\sigma}}}\,
{{\delta\, {\S_{\hbox{b}}}}\over{\delta\, g_{\mu\nu}}} & &
\nonumber\\ 
+ \, {1\over\sqrt{g}}\, \left[{1\over{D-2}}\, 
\left(g_{\mu\nu}\, {{\delta\, {\S_{\hbox{b}}}}\over{\delta\, g_{\mu\nu}}}\right)^2
-\, {{\delta\, {\S_{\hbox{b}}}}\over{\delta\, g_{\mu\nu}}}\, 
{{\delta\, {\S_{\hbox{b}}}}\over{\delta\, g^{\mu\nu}}}
-\, {1\over 2}\, {{\delta\, {\S_{\hbox{b}}}}\over{\delta\, \phi^I}}\,
{{\delta\, {\S_{\hbox{b}}}}\over{\delta\, \phi^I}}\right] = & &  
\nonumber\\
\, \sqrt{g}\, \left[ V \left(\phi\right)\, -\, 
{1\over 2}\, \nabla_\mu\, \phi^I\, \nabla^\mu\, \phi^I 
+\, \R + 2\, \Lambda + a_o\, \R^2 + b_o\, \R_{\mu\nu}\, \R^{\mu\nu} 
+\, c_o\, \R_{\mu\nu\rho\sigma}\, \R^{\mu\nu\rho\sigma}\right] . & &
\label{cl}  
\eeqa
As advertised earlier, the $\alpha'$ corrections from the bulk, in the form of higher-curvature
terms, modifies the HJ constraint. This HJ constraint will play a central role in the 
remainder. Indeed, the bulk/boundary correspondence proposes to replace the bulk action 
${\S_{\hbox{b}}}$ in (\ref{cl}), induced on the timelike foliations, with that of an effective 
$(D-1)-$dimensional boundary theory. It is easy to see then that the HJ constraint in (\ref{cl}) 
allows us to determine the coefficients of all the local terms in that boundary action, which will
in effect include contributions from the higher-curvature corrections. In
relation with the higher-curvature corrections, we should also notice that
they do induce in the boundary Lagrangian, quartic powers of the extrinsic
curvature $K$, schematically  denotes as $K^4$, besides the quadratic terms
$K^2$. As a consequence, the Hamiltonian $\H_T$ in (\ref{cd})  must also
include terms that are quartic in the conjugate momentum $\pi^{\mu\nu}$, such
as as $\pi^4$.  The reason, we chose not to include the $\pi^4$ terms in
$\H_T$, and focus only on the corrections coming  from the quadratic terms in
$K$, is that, it is in principle possible to generate the quartic terms such
as  $K^4$, in a Wilsonian manner\footnote{This way of viewing the
higher-curvature corrections was suggested to us  by Herman Verlinde.}, as
effective interactions. This is done by integrating out some very heavy
auxiliary field  $\chi$, with mass much higher than the cut-off scale in the
boundary theory, and which enters the boundary Lagrangian  in the form
$M_{\chi}^2\, \chi^2 + \xi\, \chi\, K^2$.  
\smallskip

Finally, using the definition of the extrinsic curvature given by equation (\ref{zj}) of the appendix, 
the radial flow of $g_{\mu\nu}$ follows straightforwardly from the expression of the canonical momentum 
$\pi^{\mu\nu}$ in equation (\ref{ch}), and it is found to be
\beqa
K_{\mu\nu} & = & {1\over 2}\, \L_r\, g_{\mu\nu} = -\, {1\over 2}\, {{\dot{g}}_{\mu\nu}} =
\left(\pi_{\mu\nu} - {\pi\over{D-2}}\, g_{\mu\nu}\right)
\, +\, \left(2\, a_o + {{4\, c_o}\over{D-2}}\right)\, {{\R\, \pi}\over{D-2}}\, g_{\mu\nu}
\nonumber\\
& & \qquad\qquad\qquad\qquad\qquad
+\, \left(b_o - 4\, c_o\right)\, 
{{\R_{\rho\sigma}\, \pi^{\rho\sigma}}\over{D-2}}\, g_{\mu\nu}
-\, 2\, a_o\, \R\, \pi_{\mu\nu}\, +\, \left(b_o - 4\, c_o\right)\, 
{{\R_{\mu\nu}\, \pi}\over{D-2}}
\nonumber\\
& & \qquad\qquad\qquad\qquad\qquad
-\, b_o\, \left({\R_\mu}^\rho \, \pi_{\rho\nu}\, +\, 
{\R_\nu}^\rho \, \pi_{\rho\mu}\right)\, +\, 4\, c_o\, \R_{\mu\rho\nu\sigma}\, \pi^{\rho\sigma}\, ,
\label{cj}\\
K & = & {K^\mu}_\mu = -\, {\pi\over{D-2}}\, +\,
\left(2\, a_o + b_o + {{4\, c_o}\over{D-2}}\right)\, {{\R\, \pi}\over{D-2}}\,
+\, \left(3\, b_o - D\, b_o - 4\, c_o\right)\, {{\R_{\mu\nu}\, \pi^{\mu\nu}}\over{D-2}}\, .
\label{ck}
\eeqa
Therefore, given the functional form of the boundary action ${\S_{\hbox{b}}}$ at slice $r= r_o$, and 
using the first-order equations (\ref{cg}) and (\ref{cj}), one can unambiguously compute the radial 
evolution of the couplings $\phi^I$ and the metric $g_{\mu\nu}$ in terms of their values on that slice. 
\smallskip
           
\section{$1/N$ corrections as a WKB approximation}

We have seen in the section~(1) that the ${AdS_D}/{CFT_{D-1}}$ correspondence involves both the large 't Hooft 
coupling $g_{YM}^2\, N >> 1$, and the large $N>>1$ limit. Because of the relation ${R^2}/\alpha' = \sqrt{g_{YM}^2 \, N}$,
relaxing the limit $g_{YM}^2\, N >> 1$ on the 't Hooft coupling reduces simply to the problem of incorporating
the $\alpha'/R^2$ corrections, as we have seen in section~(2). This was carried out systematically, by 
considering the effects of the bulk higher-curvature terms on the boundary theory. When we turn to the large 
$N>>1$ limit, the derivation of the $1/N$ corrections to the HJ constraint does not unfortunately enjoy the same 
degree of simplicity. What we seem to be missing here is a systematic method, analogous to the $\alpha'/R^2$ 
corrections case, where the $1/N$ corrections could be, for example, derived from first principles such as the 
open/closed string duality relation proposed in [30]. In the absence of such
systematic methods, our derivation  of the $1/N$ corrections relies simply on
our experience and intuition based on similar problems in other physical 
examples. One such (well known) example is the problem we face when we make
the transition from the {\it classical} HJ equation to the {\it quantum}
Schr{\"{o}}dinger equation. To see this, we recall from quantum mechanics that
the  wave amplitude to be associated with the mechanical motion of a particle
of mass $m$ have  the form \beq \psi = \psi_o\, e^{{i\over{\hbar}}\, S}\, ,
\label{da}
\eeq
where $\psi$ satisfies the Shr{\"{o}}dinger wave equation
\beq
{{\hbar^2}\over{2\, m}}\, {{\bf \nabla}^2}\, \psi - V\, \psi = 
{{\hbar}\over{i}}\, {{\partial\, \psi}\over{\partial\, t}}\, .
\label{db}
\eeq
In terms of the action $S$ the Shr{\"{o}}dinger equation can be written as
\beq
\left[ {1\over{2\, m}}\, \left({\bf \nabla}\, S\right)^2 + V\right] + 
{{\partial\, S}\over{\partial\, t}} = {{i\, \hbar}\over{2\, m}}\, {\bf \nabla}^2\, S\, .
\label{dc}
\eeq
The last equation may be called the quantum-mechanical HJ equation; it reduces to the
classical HJ equation in the limit as $\hbar$, and therefore the Compton wavelength of
the particle, goes to zero. Indeed, one is to note that ${\bf \nabla}^2\, S$ arises in association with 
$\left({\bf \nabla}\, S\right)^2$ in the evaluation of ${\bf \nabla}^2\, \psi$ in the quantum-mechanical wave 
equation. Therefore, (\ref{dc}) would be the classical HJ equation if 
$\hbar\, {\bf \nabla}^2\, S << \left({\bf \nabla}\, S\right)^2$, or, equivalently, if 
$\lambda/{2\pi} <<  p/\left({\bf \nabla \cdot p}\right)$. 
\smallskip

It is clear from the discussion above that the key element in the transition from the classical 
to the quantum HJ equation is the relation between the wave-function $\psi$
and the action given by  (\ref{da}), and the Shr{\"{o}}dinger wave equation
(\ref{db}) describing the propagation of $\psi$. It is exactly the analog of
these relations that we would need in the ${AdS_D}/{CFT_{D-1}}$
correspondence,  to be able to derive the $1/N$ corrections to the HJ
constraint in a systematic way,  and which we do not have \footnote{In [15],
it was suggested that the HJ constraint can be considered as the classical
limit of the quantum Wheeler-De Witt equation, which when written as
$e^{{i\over{\hbar}}\, S}$, contains an additional term proportional to a
second order variation of the action $S$.}. Despite this difficulty, one can
still use the above analogy, in particular the quantum-mechanical HJ equation
in (\ref{dc}), to discuss the $1/N$ corrections. The analogy becomes even more
clear if we think of $1/N$ as $\sqrt{\hbar}$. With this in mind, it is natural
to write down the following equation for the HJ constraint   
\beqa  2\, a\,
{\R\over\sqrt{g}}\,  \left[{{\delta\, {\S_{\hbox{b}}}}\over{\delta\,
g_{\mu\nu}}}\,  {{\delta\, {\S_{\hbox{b}}}}\over{\delta\, g^{\mu\nu}}} -\,
{1\over{D-2}}\, \left(g_{\mu\nu}\,  {{\delta\, {\S_{\hbox{b}}}}\over{\delta\,
g_{\mu\nu}}}\right)^2\right]  +\, 2\, b\, {\R_{\mu\nu}\over\sqrt{g}}\,
\left[{{\delta\, {\S_{\hbox{b}}}}\over{\delta\, {{g_\mu}^\rho}}}\, 
{{\delta\, {\S_{\hbox{b}}}}\over{\delta\, g_{\rho\nu}}}
-\, {{g_{\rho\sigma}}\over{D-2}}\, {{\delta\, {\S_{\hbox{b}}}}\over{\delta\, g_{\rho\sigma}}}\,
{{\delta\, {\S_{\hbox{b}}}}\over{\delta\, g_{\mu\nu}}}\right] & &
\nonumber\\
-\, 4\, c\, {\R_{\mu\nu\rho\sigma}\over\sqrt{g}}\, 
{{\delta\, {\S_{\hbox{b}}}}\over{\delta\, g_{\mu\rho}}}\,
{{\delta\, {\S_{\hbox{b}}}}\over{\delta\, g_{\nu\sigma}}}
-\, {{4\, c}\over{\left(D-2\right)^2}}\, {\R\over\sqrt{g}}\,
\left(g_{\mu\nu}\, {{\delta\, {\S_{\hbox{b}}}}\over{\delta\, g_{\mu\nu}}}\right)^2
+\, {{8\, c}\over{D-2}}\, {\R_{\mu\nu}\over\sqrt{g}}\,
g_{\rho\sigma}\, {{\delta\, {\S_{\hbox{b}}}}\over{\delta\, g_{\rho\sigma}}}\,
{{\delta\, {\S_{\hbox{b}}}}\over{\delta\, g_{\mu\nu}}} & &
\nonumber\\ 
+\, {1\over\sqrt{g}}\,
\left[{e_1}\, {{\delta^2\, S_{\hbox{b}}}\over{\delta\, g_{\mu\nu}\, \delta\, g^{\mu\nu}}}
+\, {e_2}\, g_{\rho\sigma}\, {\delta\over{\delta\, g_{\rho\sigma}}}\, 
g^{\mu\nu}\, {{\delta\, \S_{\hbox{b}}}\over{\delta\, g^{\mu\nu}}}
+\, {e_3}\, {{\delta^2\, \S_{\hbox{b}}}\over{\delta\, \phi^I\, \delta\, \phi^I}}\right] 
\nonumber\\
+ \, {1\over\sqrt{g}}\, \left[{1\over{D-2}}\, 
\left(g_{\mu\nu}\, {{\delta\, {\S_{\hbox{b}}}}\over{\delta\, g_{\mu\nu}}}\right)^2
-\, {{\delta\, {\S_{\hbox{b}}}}\over{\delta\, g_{\mu\nu}}}\, 
{{\delta\, {\S_{\hbox{b}}}}\over{\delta\, g^{\mu\nu}}}
-\, {1\over 2}\, {{\delta\, {\S_{\hbox{b}}}}\over{\delta\, \phi^I}}\,
{{\delta\, {\S_{\hbox{b}}}}\over{\delta\, \phi^I}}\right] = & &  
\nonumber\\
 \, \sqrt{g}\, \left[ V \left(\phi\right)\, -\, 
{1\over 2}\, \nabla_\mu\, \phi^I\, \nabla^\mu\, \phi^I 
+\, \R + 2\, \Lambda + a\, \R^2 + b\, \R_{\mu\nu}\, \R^{\mu\nu} 
+\, c\, \R_{\mu\nu\rho\sigma}\, \R^{\mu\nu\rho\sigma}\right] , & &
\label{dg}  
\eeqa
where $e_1$, $e_2$ and $e_3$ are the coefficients parameterizing the $1/N$ corrections to the HJ constraint
in the same way that $a_o$, $b_o$ and $c_o$ parameterize the $\alpha'$ corrections in (\ref{cl}). In fact, 
in writing down the HJ constraint (\ref{dg}), we have replaced $a_o$, $b_o$ and $c_o$ by the new coefficients
$a$, $b$ and $c$ allowing the latter to include extra $1/N$ contributions besides the $\alpha'$ corrections.
Therefore, we can write
\beq
a = a_o + a_1\, ,\, 
b = b_o + b_1\, ,\,
c = c_o + c_1\, ,
\label{dh}
\eeq
where $a_1$, $b_1$ and $c_1$ are taken to parameterize the $1/N$ corrections.
\smallskip

In this paper, we take the point of view that equation (\ref{dh}) represent the correct HJ 
constraint taking into account the leading order corrections in $\alpha'$ and $1/N$. Using 
(\ref{dh}), we shall determine in the next section the various bulk/boundary relations that
follow from it.
\smallskip

\section{The local boundary action terms revisited} 

One of the remarkable features of the ${AdS_D}/{CFT_{D-1}}$ correspondence is that the bulk/boundary correspondence 
is captured by the HJ constraint in equation (\ref{dg}). The latter has, in particular, the advantage of containing 
both the $\alpha'$ and $1/N$ corrections (in the leading order). Extending, therefore, previous work on the $RG-$flow 
beyond the low-energy, strong coupling, large $N$ limit. It is also important to realize that the HJ constraint in 
(\ref{dg}) proposes that we replace the bulk action $S_{\hbox{b}}$, induced on the timelike slice due the foliation 
of the bulk spacetime, with that of an unknown effective $(D-1)-$dimensional boundary theory, whose action we denote by 
action $S^{\hbox{eff}}$. With this in view, the HJ constraint plays now the role of a functional differential equation 
allowing for the determination of the functional form of the local terms in the boundary action $S^{\hbox{eff}}$, as we
shall see below. 
\smallskip

It is well know that the $RG-$flow of quantum field theory in a curved background induces, in the 
effective action, an Einstein gravity term plus a cosmological constant. Indeed, a computation of 
the $\left\langle\, T_{\mu\nu}\, \right\rangle$ for the quantum field and its subsequent regularization 
is found to renormalize both the Einstein tensor and the cosmological constant. Therefore, at the cut-off 
scale $\mu$, a general form for the effective action, $S^{\hbox{eff}}$, is given by
\beq
\S^{\hbox{eff}} \left(g,\phi\right)  =  
\S_{\hbox{l}} \left(g,\phi\right) + \S_{\hbox{nl}} \left(g,\phi\right)\, ,
\label{eaa}
\eeq
where $\S_{\hbox{l}}$ represent the local part of the effective action whose form is
\beqa
\S_{\hbox{l}} \left(g,\phi\right) & = &
\int_{D-1}\, \sqrt{g}\, {d^{D-1}}x\,
\left[\kappa\left(\phi\right)\, \left(\R - 
{1\over 2}\, \nabla^\mu\, \phi^I\, \nabla_\mu\, \phi^I\right) + 
U \left(\phi\right) \right.
\nonumber\\
& & \qquad\qquad \left.
- \A \left(\phi\right)\, \R^2 
- \B \left(\phi\right)\, \R^{\mu\nu}\, \R_{\mu\nu}
- \C \left(\phi\right)\, \R^{\mu\nu\rho\sigma}\, \R_{\mu\nu\rho\sigma}\right]\, ,
\label{eab}
\eeqa
where $\R$, $\R^2$, $\R^{\mu\nu}\, \R_{\mu\nu}$ and $\R^{\mu\nu\rho\sigma}\, \R_{\mu\nu\rho\sigma}$ denote the 
$(D-1)-$dimensional curvature terms constructed form the boundary metric $g_{\mu\nu}$ in (\ref{cc}). The
boundary values of the scalar fields $\phi^I$ are to be equated with the dimensionless coupling constants 
of the boundary theory, and $U\left(\phi\right)$, $\kappa\left(\phi\right)$, $\A\left(\phi\right)$,
$\B\left(\phi\right)$ and $\C\left(\phi\right)$ are local functions of these 
couplings. $\S_{\hbox{nl}}$ contains, on the other hand, all higher derivative and non-local terms subject to 
the symmetries inherited from the bulk\footnote{The usual quartic, quadratic and logarithmic divergences for 
quantum fields coupled to curved spacetime are contained in the local action $\S_{\hbox{l}}$ through $U$, $\kappa$ 
and $(\A, \B, \C)$, respectively. The non-local action $\S_{\hbox{nl}}$ may also contain extra logarithmic divergences.}. 
In terms of the non-local action $\S_{\hbox{nl}}$, the boundary theory operators $\left\langle \O_I\right\rangle$ and 
energy-momentum tensor $\left\langle T_{\mu\nu}\right\rangle$, are given by
\beqa
{1\over{\sqrt{g}}}\, {{\delta \S_{\hbox{nl}}}\over{\delta \phi^I}} & \equiv &
\left\langle \O_I\right\rangle \, ,
\label{eba}\\
{1\over{\sqrt{g}}}\, {{\delta \S_{\hbox{nl}}}\over{\delta g^{\mu\nu}}} & \equiv &
\left\langle T_{\mu\nu}\right\rangle\, .
\label{ebb}
\eeqa
\smallskip

Our goal now is to determine the local boundary terms in $\S^{\hbox{eff}}$. For this, we need to 
insert the effective action $\S^{\hbox{eff}}$ into the HJ constraint (\ref{dg}), equating 
contributions from the left hand side with terms on the right hand side that have the same 
functional form. By treating the metric $g_{\mu\nu}$ and the scalars $\phi^I$ as arbitrary 
classical fields, this procedure generates a set of bulk/boundary relations for the unknown 
functions in the local action $\S_{\hbox{l}}$, which are  
\beqa
2\, \Lambda + V & = & 
\left[{1\over 4}\,  {{D-1}\over{D-2}}\, {U^2} - {1\over 2}\, \left(\partial_I\, U\right)^2 \right]
\nonumber\\
& &
+\left[-\, {{e_1}\over 4}\, \left({D^2}-1\right)\, U 
- {{e_2}\over 4}\, {\left(D-1\right)^2}\, U
+ {e_3}\, \partial^I\, \partial_I\, U \right]\, ,
\label{ec}\\
& &\nonumber\\& &\nonumber\\
1 & = &
\left[{1\over 2}\, {{D-3}\over{D-2}}\, \kappa\, U - \partial^I\, \kappa\, \partial_I\, U\right]
+ \left[-\, {a\over 2}\, {{D-1}\over{D-2}}\, U^2 - {b\over 2}\, {1\over{D-2}}\, U^2
- c\, {1\over{\left(D-1\right)^2}}\, U^2\right]
\nonumber\\
& & 
+ \left[-\, {{e_1}\over 4}\, {\left({D^2} - 5\right)}\, \kappa 
- {{e_2}\over 4}\, {\left(D-3\right)^2}\, \kappa
+ {e_3}\, \partial^I\, \partial_I\, U \right]\, , 
\label{ed}\\
& &\nonumber\\& &\nonumber\\
a & = &
\left[{1\over 4}\, {{D-1}\over{D-2}}\, \kappa^2 - {1\over 2}\, {{D-5}\over{D-2}}\, U\, \A
- {1\over 2}\, \partial^I\, \kappa\, \partial_I\, \kappa + \partial^I\, U\, \partial_I\, \A\right]
\nonumber\\
& &
+ \left[-\, a\, {{D-3}\over{D-2}}\, \kappa\, U 
- 2\, c\, {{\left(D-1\right)\,\left(D-3\right)}\over{\left(D-2\right)^2}}\, \kappa\, U\right]
\nonumber\\
& &
+ \left[{e_1}\, \left({{{D^2}-9}\over 4}\, \A + \B\right) 
+ {{e_2}\over 4}\, {\left(D-5\right)^2}\, \A
- {e_3}\, \partial^I\, \partial_I\, \A\right]\, ,
\label{ef}\\
& &\nonumber\\& &\nonumber\\
b & = & 
\left[- \kappa^2 - {1\over 2}\, {{D-5}\over{D-2}}\, U\, \B + \partial^I\, U\,  \partial_I\, \B\right]
+ \left[-\, b\, {{D-3}\over{D-2}}\, \kappa\, U - 4\, c\, \kappa\, U\right]
\nonumber\\
& &
+ \left[{e_1}\, \left({{{D^2}-5}\over 4}\, \B + 2\, \A + 4\, \C\right) 
+ {{e_2}\over 4}\, {\left(D-5\right)^2}\, \B
- {e_3}\, \partial^I\, \partial_I\, \B \right]\, ,
\label{eg}\\
& &\nonumber\\& &\nonumber\\
c & = &
\left[-\, {1\over 2}\, {{D-9}\over{D-2}}\, U\, \C + \partial^I\, U\, \partial_I\, \C\right]
+ \left[{{e_1}\over 4}\, \left({D^2} - 17\right)\, \C 
+ {{e_2}\over 4}\, {\left(D-9\right)^2}\, \C\right]\, ,
\label{eh}\\
& &\nonumber\\& &\nonumber\\
\beta^I\, \partial_I\, \kappa & = & 
\left[-\, \left(D-1\right)\, \kappa + {{2\, \left(D-2\right)}\over U}\right]
\nonumber\\
& + &
 \left[{{e_1}\over 2}\, {{\left({D^2}-1\right)\, \left(D-2\right)}\over U}\, \kappa
+ {{e_2}\over 2}\, {{\left(D-2\right)\, \left(D-1\right)^2}\over U}\, \kappa
- 2\, {e_3}\, {{\left(D-2\right)}\over U}\, \partial^I\, \partial_I\, \kappa\right],
\label{ei}
\eeqa 
where the beta-functions $\beta^I$'s are defined by
\beq  
\beta^I \left(\phi\right) = -\, 2\, {{D-2}\over U}\, \partial_I\, U\, .
\label{ej}
\eeq
In addition, we have to this order in the expansion, terms involving the functional derivatives of 
the non-local action $S_{\hbox{nl}}$. The bulk/boundary relations for them are
\beqa
\left\langle {T^\mu}_\mu \right\rangle \equiv \left\langle T \right\rangle & = &
{{\beta^I}\over 2}\, \left\langle \O_I \right\rangle\, ,
\label{ek}\\
& &\nonumber\\& &\nonumber\\
V & = & 
\left[ {{{\left\langle T\right\rangle }^2}\over{D-2}} 
-  \left\langle T^{\mu\nu}\right\rangle\, \left\langle T_{\mu\nu}\right\rangle
- {1\over 2}\, \left\langle \O^I\right\rangle\, \left\langle \O_I\right\rangle\right]
\nonumber\\
& &
+ \left[{e_1}\, \left\langle T^{\mu\nu}\, T_{\mu\nu}\right\rangle 
+ {e_2}\, \left\langle T^2 \right\rangle
+ {e_3}\, \left\langle \O^I\, \O_I\right\rangle \right]\, .
\label{el}
\eeqa
In the next section, we shall use these new bulk/boundary relations to study the cosmological constant 
problem. In particular, we are interested to see whether the solution proposed
in [15], for the vanishing  of $\Lambda$, continues to hold in the presence
of the $\alpha'$ and the $1/N$ corrections.   
\smallskip

\section{What is new on the cosmological constant problem?}

The problem of the cosmological constant is why the vacuum energy density is zero 
or extremely small by particle physics standards. It is a hard problem 
because it involves not only the high-energy but the low-energy physics as well. 
It is not sufficient, for example, to find a cosmological constant that is zero at 
high energies (near the Planck scale), one must also explain the absence of the vacuum 
contributions as the scales run to low energies. This low-energy aspect of the 
cosmological constant is, in fact, the most puzzling, and seems to require some 
fundamental new ideas in the basic principles of low-energy effective field theories, 
$RG-$flow and gravity. But the low-energy physics in the standard framework of $4-$dimensional 
effective field theory does not seem to offer a solution to the problem\footnote{For a complete 
review on these issues see the paper by Weinberg in [22].}. On the other
hand, it is very hard  to change the low-energy theory in a sensible way,
given all of the well known theoretical and  experimental success. Faced with
this riddle, one way out would be to imagine a scenario in  which the
observed $4-$dimensional universe, where the problem is severely posed, is
related  to a world of a higher dimension. If the higher-dimensional world
does not obey the usual assumptions  of $4-$dimensional low-energy effective
field theories, which lead to the cosmological constant problem,  one may then
find a solution to this problem within in this scenario. 
\smallskip

In the following we will reexamine the cosmological constant problem using a scenario in which
the observed $4-$dimensional universe is embedded into a higher-dimensional background of dimension
$D = 5$. Our approach is directly motivated by the new insights from string theory through the 
$AdS_5/CFT_4$ correspondence, as well as by recent ideas that have appeared in the study of warped 
string compactification scenarios along the lines of Randall and
Sundrum\footnote{Despite recent attempts  in [15], it does not exist yet a
complete and consistent embedding of the Randall-Sundrum  scenario within
string or M-theory.} in [21,15,22,16], reviving earlier work by Rubakov and
Shaposhnikov [22].  The starting point of our discussion is the holographic
formulation of the $RG$ equations in which the $RG$  scale is treated as a
physical extra dimension. We also assume the warp geometry for the
$5-$dimensional bulk  spacetime\footnote{Such backgrounds could be obtained,
for example, via F-theory compactification on Calabi-Yau  fourfolds [31].},
which generalizes the $AdS_5/CFT_4$ duality to $4-$dimensional boundary
theories with dynamical gravity, as our world. Following [15], and applying
the results of section~(5) to a $5-$dimensional bulk spacetime of warp
geometry and $\Lambda = 0$, one finds that the HJ evolution equations  in the
bulk can also be reformulated as an $RG-$flow equations\footnote{To find the
$RG-$flow equations of the  boundary effective theory, one solves for the
evolution equations in (\ref{cj}) and (\ref{cg}) using the warp  geometry
ansatz for the bulk, after replacing by the constraints from the HJ constraint
in (\ref{dg}).} for the $4-$dimensional boundary effective action, even after
the inclusion of the $\alpha'$ and $1/N$ corrections. Our calculations,
therefore, extends previous results found within the context of classical
$5-$dimensional  supergravity [15], and thus within the large $N$ and large 
`t Hooft coupling limit, to the regime where these  limits are relaxed. In
particular, new interesting bulk/boundary relations were found, suggesting an
intimate  connection between the $RG-$flow symmetry of the boundary effective
action and the bulk Einstein's equations. 
\smallskip

Let us now address the consequences of the $RG-$flow symmetry of the boundary effective action,
in the presence of the leading order corrections in $\alpha'$ and $1/N$, on the $4-$dimensional 
cosmological constant. Using the same line of reasoning as in [15], our
$RG-$flow equations imply  also that once we have a solution for the
gravitational part of the boundary effective action at  one scale, there is a
solution along the whole $RG-$trajectory. As a result, assuming that the 
boundary cosmological constant is canceled at high energies (due to extended
supersymmetry,  for example), it will naturally remain zero under the
$RG-$flow. So it appears as if the boundary  cosmological constant continues
to decouple from the $RG-$induced vacuum energy of the matter  fluctuations,
even after relaxing the large $N$ and the large 't Hooft coupling limit. As we
will show now, this decoupling arises due to a cancellation between the
contraction rate of  the warp factor and any variation in the matter induced
vacuum energy, in close similarity with with the mechanism proposed in [15].
Using a $5-$dimensional background of warp geometry with vanishing  $\Lambda$
as our bulk spacetime, the field equations that follow from the effective
action (\ref{eaa})  are then the $4-$dimensional Einstein equation and the
scalar field equations \beqa \kappa\, \left(\R_{\mu\nu} - {1\over 2}\,
g_{\mu\nu}\, R\right) - {1\over 2}\, g_{\mu\nu}\, U\left(\phi\right) & = & 
\left(\A\, {}^{(1)} H_{\mu\nu} + \B\, {}^{(2)} H_{\mu\nu} + \C\,
H_{\mu\nu}\right)  \nonumber\\ && +\, \T^{\phi}_{\mu\nu} \left(\kappa, \A, \B,
\C, \phi, g_{\mu\nu}\right)  - \left\langle T_{\mu\nu}\right\rangle\, ,
\label{fa} \eeqa
\beqa
 \nabla_{\mu}\, \left(\kappa\, \nabla^{\mu}\, \phi^I\right) + \partial_I\, \kappa\, 
\left(R - {1\over 2}\, \nabla^{\lambda}\, \phi^I\, \nabla_{\lambda}\, \phi^I\right)
& = & \left(\partial_I\, \A\, \R^2 + \partial_I\, \B\, \R^{\mu\nu}\, \R_{\mu\nu}
+ \partial_I\, \C\, \R^{\mu\nu\rho\sigma}\, \R_{\mu\nu\rho\sigma}\right)
\nonumber\\
&& -\, \partial_I\, U - \left\langle \O_I\right\rangle\, ,
\label{fb}
\eeqa
where ${}^{(1)} H_{\mu\nu}$, ${}^{(2)} H_{\mu\nu}$ and $H_{\mu\nu}$ are the contributions
to the field equations from the higher curvature terms, and are given by
\beqa
{}^{(1)} H_{\mu\nu} & = & 2\, \nabla_\mu\, \nabla_\nu\, \R - 2\, g_{\mu\nu}\, \Box\, \R
- {1\over 2}\, g_{\mu\nu}\, \R^2 + 2\, \R\, \R_{\mu\nu}\, ,
\label{fca}\\
{}^{(2)} H_{\mu\nu} & = & 2\, \nabla_\alpha\, \nabla_\nu\, {\R^\alpha}_\mu
- \Box\, \R_{\mu\nu} - {1\over 2}\, g_{\mu\nu}\, \Box\, \R 
+ 2\, {\R_\mu}^\alpha\, \R_{\alpha\nu} 
- {1\over 2}\, g_{\mu\nu}\, \R^{\alpha\beta}\, \R_{\alpha\beta}\, ,
\label{fcb}\\
H_{\mu\nu} & = & 2\, \nabla_\mu\, \nabla_\nu\, \R - 4\, \Box\, \R_{\mu\nu} + 
2\, \R_{\mu\alpha\beta\gamma}\, {\R_\nu}^{\alpha\beta\gamma}
- {1\over 2}\, g_{\mu\nu}\, \R^{\alpha\beta\gamma\delta}\, \R_{\alpha\beta\gamma\delta}
- 4\, \R_{\mu\alpha}\, {\R^\alpha}_\nu
\nonumber\\
& &  +\, 4\, \R^{\alpha\beta}\, \R_{\alpha\mu\beta\nu}\, .
\label{fcc}
\eeqa
$\T^{\phi}_{\mu\nu}$ represents the stress energy-momentum tensor of the scalar 
fields $\phi^I$. Besides $\phi^I$, $\T^{\phi}_{\mu\nu}$ depends also on the functions 
$\kappa$, $\A$, $\B$, $\C$, their covariant derivatives and the various curvature terms 
of the metric $g_{\mu\nu}$. $\left\langle T_{\mu\nu}\right\rangle$ and $\left\langle \O_I\right\rangle$
were defined earlier in (\ref{eba}) and (\ref{ebb}), and they represent the boundary expectation values 
to which the metric $g_{\mu\nu}$ and the scalar fields $\phi^I$ couple, respectively. 
\smallskip

At this point, one could make use of the $RG-$flow equations of the boundary effective theory to deduce the 
$RG-$trajectories of all the quantities appearing in the field equations (\ref{fa}) and (\ref{fb}), and show
the decoupling mechanism that is claimed to arise for the cosmological constant. Since this approach 
has already been used in the previous literature such as in [15], what we
propose here is a much simpler and  direct method making use of the
bulk/boundary relations derived in section~(5). To address the consequences of
 the bulk/boundary relations on the boundary cosmological constant, let us
take the trace of the $4-$dimensional Einstein's equations in (\ref{fa}),
yielding \beq \kappa\, R = \left\langle T\right\rangle - 2\, U 
+ 2\, \left(3\, \A + \B + \C\right) \Box \R = \kappa\, \Lambda^{(4)}\, ,
\label{fd}
\eeq
where we have assumed the boundary theory to be at an energy scale much less than the cut-off scale $\mu$, 
so that the scalar fields are practically independent of the $4-$dimensional boundary coordinates, {\it i.e.,} 
$\nabla_\mu\, \phi^I \left( x\right) = 0$. Clearly, the terms on the right-hand side of (\ref{fd}) represents
the effective cosmological constant on the boundary. We would have $\Lambda^{(4)} = 0$ if the first two terms 
on the right-hand side of (\ref{fd}) cancel each other, and the third term is zero. First how do we make the
third term vanish? Since the HJ constraint, and the hence the bulk/boundary relations derived from it, are 
nothing more than constraints on the variations of both $\S_{\hbox{l}}$, and $\S_{\hbox{nl}}$ in $\S^{\hbox{eff}}$, 
one may consider these constraints for any boundary field configuration, including a preferred one,
such that $3\, \A + \B + \C = 0$. Using this condition, the trace of the Einstein equation in (\ref{fd}) becomes 
\beq
\kappa\, R = \left\langle T\right\rangle - 2\, U = \kappa\, \Lambda^{(4)}\, .
\label{fe}
\eeq
The condition $3\, \A + \B + \C = 0$ is easily seen to be satisfied if the higher-curvature contributions entered 
the local effective $S_{\hbox{l}}$ in (\ref{eab}) as a Gauss-Bonnet term 
$\left(\R^2 - 4\, \R^{\mu\nu}\, \R_{\mu\nu} + \R^{\mu\nu\rho\sigma}\, \R_{\mu\nu\rho\sigma}\right)$.
This Gauss-Bonnet term was considered in [16] and [18] in the study of naked
singularities within the context of brane world scenarios . This is not the
point of view we take here. We consider,  instead, the situation where the
condition $3\, \A + \B + \C = 0$ is satisfied for arbitrary coefficients $\A$,
$\B$ and $\C$. But since the Gauss-Bonnet term is a topological invariant on
the $4-$dimensional boundary,  only two of them are independent , so me may
choose $\C = 0$. The coefficients $\A$ and $\B$ satisfy then the  condition
$3\, \A + \B = 0$. Using the bulk/boundary relations (\ref{ef}), (\ref{eg})
and (\ref{eh}), the  conditions $\C = 0$ and $3\,\A + \B = 0$ translate in
thus into conditions on the bulk parameters $\left(a,b,c\right)$, where $c =
0$ and $a$ and $b$ being related to each other.  
\smallskip

Now, let us turn to the remaining two terms on the right-hand side of (\ref{fe}). At first sight it is not obvious 
why should $\left\langle T\right\rangle$ and $2\, U$ cancel each other. However, by invoking again the fact that 
the HJ constraint is simply a condition on the variations of $\S_{\hbox{l}}$ and $\S_{\hbox{nl}}$ which hold for 
an arbitrary field configuration, one may consider it for a flat boundary spacetime with constant scalars. 
In this case, using the bulk/boundary relations in (\ref{ec}) and (\ref{el}), after setting $D=5$ and $\Lambda = 0$, 
we find that $\left\langle T\right\rangle$ and $2\, U$ are given by the following expressions
\beqa
\left(2\, U\right)^2 & = & 12\, V + 6\, \left(\partial_I U\right)^2 + 24\, \left( 3\, e_1 + 2\, e_2 \right)\, U
-12\, e_3 \, \partial^I\, \partial_I U\, ,   
\label{ffa}\\
\left\langle T\right\rangle^2 & = & 12\, V + 6\,  \left\langle \O_I\right\rangle^2 
- 3\, \left(e_1 + 4\, e_2 \right)\, \left\langle T^2 \right\rangle - 
12\, e_3\, \left\langle \O_I^2 \right\rangle\, .
\label{ffb}
\eeqa
So far, only the trace of the Einstein equation in (\ref{fa}) and the identities (\ref{ffa}) and (\ref{ffb}) 
(from the HJ constraint) did enter our analysis of the boundary cosmological constant. To progress further
we make of the equation of motion fro $\phi^I$, which for $\nabla_\mu\,  \phi^I = 0$ and flat boundary 
spacetime reads
\beq
\partial_I\, U + \left\langle \O_I\right\rangle = 0\, .
\label{fg}
\eeq
Now, inserting (\ref{fg}) into both $\left(2\, U\right)^2$ and 
$\left\langle T\right\rangle^2$, and evaluating their difference afterwards, we
find 
\beq
\left(2\, U\right)^2 - \left\langle T\right\rangle^2 =
12\, e_3\, \left[\left\langle \O_I^2 \right\rangle - \partial^I\, \partial_I U\right]
+ 3\, e_1\, \left[24\, U + \left\langle T^2 \right\rangle\right]
+ 12\, e_2\, \left[4\, U + \left\langle T^2 \right\rangle\right]\, .
\label{fh}
\eeq
The above relation cannot be simplified further since we have already made used of all the 
equations that are available to us (which are the equations of motion and the HJ constraint). 
The consequences of this relation on the cosmological constant problem within the holographic 
$RG-$flow approach are the topic of the next section. 
\smallskip

\section{Discussion}

It appears from (\ref{fh}) that $2\, U$ and $\left\langle T\right\rangle$ would not cancel each 
other in the presence of the leading $1/N-$corrections, parametrized by the $e_1$, $e_2$ and $e_3$ 
coefficients. From equation (\ref{fe}), we see that this mismatch between $2\, U$ and 
$\left\langle T\right\rangle$ implies a non-zero effective cosmological constant $\Lambda^{(4)}$ 
on the boundary, which is in clear distinction from the results of [15]. 
In [15], since the authors were only considering the large $N$ limit, for them 
$e_1 = e_2 = e_3 = 0$, and thus they obtained the cancellation between $2\, U$ and
$\left\langle T\right\rangle$, necessary for the vanishing of the boundary cosmological constant. 
In geometric terms, this result was interpreted as meaning that there exist a natural mechanism in 
which the vacuum energy that is generated on the $4-$dimensional brane world, as we flow towards the 
$IR$, is canceled by the ever decreasing warp factor of the $5-$dimensional geometry. From the holographic 
$RG-$flow perspective (based on the HJ formalism), this result shows that, in the strong 't Hooft coupling 
and large $N$ regime, the potential energy $U$ is canceled by the trace of the stress energy tensor at all 
scales, once this achieved at one particular scale\footnote{Both interpretations hold only in the case of 
a $5-$dimensional background of warp geometry.}, yielding thus the $RG-$stability of the cosmological
constant. Given the usual difficulties in reconciling the $RG-$flow intuition and the observational 
evidence for a small cosmological constant, this is certainly a useful progress towards the final solution.
It is important to notice that the $RG-$stability of the cosmological constant, $\Lambda^{(4)}$,
established in the strong 't Hooft coupling and large $N$ regime of the boundary theory, is not 
restricted to any preferred value for $\Lambda^{(4)}$. This leaves, of course, the question of whether it is
possible to pick up naturally initial conditions in the $UV$ for which $\Lambda^{(4)} = 0$.
\smallskip

In our actual calculation, we have not address at all this question, rather what we were interested in is to 
extend the $RG-$stability of the cosmological constant to the regimes where the strong 't Hooft coupling
and the large $N$ limits are relaxed. What we found, in this case, is that the fate of the $RG-$stability mechanism,
of [15], is not sensitive to the $\alpha'-$corrections, which were introduced to account for the 
relaxation of the strong 't Hoof coupling limit. However, one sees from equation 
(\ref{fh}) that the $1/N-$corrections do seem on the other hand to ruin the $RG-$stability of the cosmological constant 
if no other equations are supplemented at this order to (\ref{fh}). As we have seen in 
section~(4), the derivation of $1/N-$corrections are less systematic and much harder to implement in 
the HJ formulation than the $\alpha'-$corrections. Using the analogy with the transition
from the HJ equation to the Schr{\"{o}}dinger equation, and treating $1/N$ as $\sqrt{\hbar}$, the
$1/N-$corrections are expressed as the second order variation of the boundary action. Although, this a 
a good starting point to probe the effects of the $1/N-$corrections, it is clear that one needs further
information and better knowledge, especially on the side of the boundary matter sector\footnote{After all it is the matter 
fields on the boundary that form representations of the boundary gauge group, which makes them sensitive to the 
choice of $N$.} to remove the arbitrariness left in the coefficients $e_1$, $e_2$ and $e_3$ parameterizing
the $1/N-$corrections. It is very plausible that when more systematic methods become 
available\footnote{According to suggestions made in [15], systematic
methods for deriving the $1/N-$corrections could be found using the non-local loop equations in [33] or
string field theory.} further relations could found between the potential $U$ and the 
boundary operators such as $\left\langle \O_I^2 \right\rangle$ and $\left\langle T^2 \right\rangle$, leading
to the cancellation among the terms on the right-hand side of (\ref{fh}). 
So instead of using equation (\ref{fh}) to declare the failure of the $RG-$stability
of the cosmological constant, outside the regime of strong 't Hooft coupling and 
large $N$ limits, we take the point of view that it calls for a better understanding of the $1/N-$corrections 
beyond the simple addition of the second order variation of the 
boundary effective action to the HJ constraint. 
\smallskip

Going now back to equation (\ref{fh}), it is very plausible just from the CFT point of view,
to have a theory where
\beqa
\left\langle \O_I^2 \right\rangle & \propto & \partial^I\, \partial_I U\, ,
\label{fia}\\
\left\langle T^2 \right\rangle & \propto & \left\langle T \right\rangle \sim U\, .
\label{fib}
\eeqa
Furthermore, in equation (\ref{dg}) since both of the coefficients $e_1$ and $e_2$
multiply the second order variation of the boundary effective action, with the respect 
to the metric $g_{\mu\nu}$, we expect that they are not independent, and hence
$e_1 \propto e_2$. Combining this relation with the relations from (\ref{fia}) and (\ref{fib}),
we see that there is much room for the right-hand side of (\ref{fh}) to vanish, allowing
us to recover the $RG-$stability of the cosmological constant in the presence of the
leading $1/N-$corrections. Hopefully, we will come back in future work to prove the additional
relations (\ref{fia}) and (\ref{fib}) needed to preserve the cancellation between the potential 
energy $U$ and the and the trace of the stress-energy tensor $\left\langle T \right\rangle$ 
in (\ref{fh}). 
\smallskip

Finally, it would be interesting to use the new bulk/boundary relations derived in 
section~(5) to study the Randall-Sundrum scenario. We treat this question in [34]. 
\smallskip

\section{Appendix}

The purpose of this section is to give a Hamiltonian formulation of the higher-curvature
theory considered in section~(3), which is represented by the $D-$dimensional bulk action
\beq
S_T = S_1 + S_2 + S_3\,,
\label{za}
\eeq
where $S_1$, $S_2$ and $S_3$ are given by 
\beqa
S_1 & = & \int_D\, \sqrt{\G}\, dr\, {d^{D-1}}x\, \left[{}^{\{D\}}R + 
2\, \Lambda + a_o\, {}^{\{D\}}R^2 + b_o\, {}^{\{D\}}R_{\mu\nu}\, {}^{\{D\}}R^{\mu\nu} \right.
\nonumber\\
& & 
\left. \qquad\qquad\qquad\qquad\qquad\qquad\qquad\qquad\qquad 
+ c_o\, {}^{\{D\}}R_{\mu\nu\rho\sigma}\, {}^{\{D\}}R^{\mu\nu\rho\sigma}\right]\, ,
\label{zba}\\
S_2 & = & 2\, \int_{D-1}\, \sqrt{g}\, {d^{D-1}}x\, \left[\, K + {{\K}}\left(K, \nabla\, K ; a_o , b_o , c_o \right)\, 
\right]\, ,
\label{zbb}\\
S_3 & = & \int_{D}\, \sqrt{\G}\, dr\, {d^{D-1}}x\, \left[\, V\left(\phi\right) - 
{1\over 2}\, \nabla_\mu\, \phi\, \nabla^\mu\, \phi\, \right]\, .
\label{zbc}
\eeqa
All the terms appearing in $S_1$, $S_2$ and $S_3$ were introduced and defined in section~(3).
\smallskip

To obtain a Hamiltonian formulation of the bulk action $S_T$, it will be useful to resort 
to the well-known technique in general relativity, which consists of slicing the 
$D-$dimensional bulk spacetime $\M$, with metric ${\G}_{AB}$, into an arbitrary foliation 
defined by the isosurfaces $\left\{\Sigma\right\}$ [32]. For the purpose of
studying of the holographic  $RG-$flow of theories induced on {\it timelike}
boundaries sitting at different locations in the radial direction of the bulk
spacetime $\M$, we choose to foliate $\M$ along timelike isosurfaces. Because
of  this, there will be some sign flips between our formulas and the ones that
we would have obtained had  we chosen a foliation along spacelike slices. So,
given that $\left(\M , \G_{AB} \right)$ is the $D-$dimensional bulk
spacetime\footnote{In general, even though $\M$ could be geometrically
different  from the pure $AdS_D$ form (due to the possible bulk-matter
stress-energy momentum tensor back-reaction),  it still has the same topology.
Allowing, therefore, the derivation of the gauge invariant correlators on  the
${CFT_{D-1}}$ boundary from ${AdS_D}$ bulk action [11].}, we can foliate it by
a family of  $\left(D-1\right)-$dimensional timelike hyper surfaces,
$\left\{\Sigma_r\right\}$, parametrized by the scalar function $r=$constant.
Thus, we can write the bulk metric $\G_{AB}$ as \beq ds^2 = \G_{AB}\, dx^A\,
dx^B = \G_{rr}\, dr^2 + 2\, \G_{r\mu}\, dr\, dx^{\mu} +  \G_{\mu\nu}\,
dx^\mu\, dx^\nu\, . \label{zc}
\eeq
Here and throughout all the paper our notational conventions will be to take
the upper case Latin letters such as $A$ and $B$ to denote the $D-$dimensional indices 
$\left(0,1,...,D-2,r\right)$ over $\M$, and the lower case Greek indices such as $\mu$ and 
$\nu$ to denote the $\left(D-1\right)-$dimensional indices $\left(0,1,...,D-2\right)$
spanning the $\Sigma_r$ hypersurface.
\smallskip

Let $r^\mu$ be a vector field on $\M$ satisfying $r^{\mu}\, {\nabla_\mu}\, r = + 1$, and 
let $n^\mu$ be the spacelike inward pointing vector fields normal to the timelike hypersurface 
with normalization $\G_{\mu\nu}\, n^{\mu}\, n^{\nu} = + 1$. By introducing the {\it lapse} function,
$N$, and {\it shift} vector, $N^\mu$, $r^\mu$ admits a decomposition in terms of its normal and tangential
components with respect to $\Sigma_r$, as follows
\beq
r^\mu = N^\mu - N\, n^\mu\, ,
\label{zd}
\eeq
where $N$ and $N^\mu$ are given by
\beqa
N & = & - r^\mu \,  n_\mu = - \left(n^\mu \, \nabla_\mu r\right)^{-1}\, ,
\label{zea}\\
N_\mu & = & g_{\mu\nu}\, r^\nu\, .
\label{zeb}
\eeqa 
In terms of these definitions the metric in (\ref{zc}) can be rewritten as
\beq
ds^2 = \left({N^2} + {N_\mu}\, {N^\mu}\right)\, dr^2 + 2\, N_{\mu}\, dx^{\mu}\, dr +
g_{\mu\nu}\, \left(x,r\right)\, dx^{\mu}\, dx^{\nu}\, ,
\label{zf}
\eeq
where the boundary metric on $\Sigma_r$ is related to the bulk metric by the formula
$g_{\mu\nu} = \G_{\mu\nu}  - n_\mu\, n_\nu$. Using the Gaussian normal coordinates, corresponding 
to the gauge choice $N^\mu = 0$ and $N = -1$, the metric in (\ref{zf}) takes on the simple form  
\beq
ds^2 = dr^2 + g_{\mu\nu}\, \left(x,r\right)\, dx^{\mu}\, dx^{\nu}\, .
\label{zg}
\eeq
\smallskip

Another concept entering the description of the bulk spacetime $\M$ in terms of its foliations 
$\left\{\Sigma_r\right\}$, is the notion of extrinsic curvature $K_{\mu\nu}$, which is defined
by 
\beqa
K_{\mu\nu} & = & {g_\mu}{}^\rho \, \nabla_\rho\, n_\nu\, ,
\nonumber\\
& = &  {1\over 2}\, {\L_r}\, g_{\mu\nu}\, .
\label{zh}
\eeqa
The meaning given to $K_{\mu\nu}$ is that it accounts for the ``bending'' of $\Sigma_r$ 
in $\M$. Finally, to obtain a Hamiltonian functional for general relativity, we need to express the
gravitaional action in (\ref{za}) in terms of the quantities 
$\left({g_{\mu\nu}} , K_{\mu\nu} ; N^\mu = 0 , N = -1\right)$, and their time and space derivatives. 
Splitting $S_T$ along the timelike foliations, we find the following Lagrangian:
\beqa
\L_T & = & \sqrt{g}\, \left[ \R + 2\, \Lambda + a_o\, \R^2 + b_o\, \R_{\mu\nu}\, \R^{\mu\nu} + 
c_o\, \R_{\mu\nu\rho\sigma}\, \R^{\mu\nu\rho\sigma} \right . 
\nonumber\\
& & 
+\, \left( {K^2} - K_{\mu\nu} \, K^{\mu\nu} \right) +
2 \, a_o \, \R \, \left( {K^2} -  K_{\mu\nu} \, K^{\mu\nu} \right) 
\nonumber\\
& &
\left.
+\, 2\, b_o\, \R_{\mu\nu}\, \left( K \, K^{\mu\nu} - K^{\mu\rho} \, {K_\rho}^\nu \right) +
4\, c_o\, \R_{\mu\nu\rho\sigma}\, K^{\mu\rho} \, K^{\nu\sigma} + \O\left({K^4} \right) \right] 
\nonumber\\
& & 
+\, \sqrt{g}\, \left[ V\left(\phi\right)  - {1\over 2}\, \nabla_\mu \phi^I \, \nabla^\mu \phi^I
- {1\over 2}\, \left(\dot{\phi^I} \right)^2 \right] \, ,
\label{zi}
\eeqa  
where $\R_{\mu\nu\rho\sigma}$, $\R_{\mu\nu}$, and $\R$ denote the $\left(D-1\right)-$dimensional 
Riemann tensor, Ricci tensor and Ricci scalar respectively. The sum over the scalar field index 
$I$ is understood in the text and hereafter. Using this Lagrangian, the canonical momenta conjugate 
to $\phi^I$ and $g_{\mu\nu}$ are 
\beqa
\pi_I & = &   {1\over\sqrt{g}}\,  {{\partial\, L_T}\over{\partial\, {\dot{\phi^I}}}} 
= -\, \dot{\phi^I} \, ,
\nonumber\\
\pi_{\mu\nu} & = & {1\over\sqrt{g}}\, {{\partial\, L_T}\over{\partial\, {\dot{g}}^{\mu\nu}}}\, 
= \left( K_{\mu\nu} - K\, g_{\mu\nu} \right) + 
2\, a_o\, \R \, \left( K_{\mu\nu} - K\, g_{\mu\nu} \right) 
-\, b_o\, \left( K\, \R_{\mu\nu} \, +\,  \R_{\rho\sigma}\, K^{\rho\sigma}\, g_{\mu\nu} \right) 
\nonumber\\
& & \qquad\qquad\qquad 
+\, b_o\, \left( {\R_\mu}^\rho \, K_{\rho\nu} \, +\, {\R_\nu}^\rho \, K_{\rho\mu} \right)
-\, 4\, c_o\, \R_{\mu\rho\nu\sigma}\, K^{\rho\sigma}\, +\, \O \left( K^3 \right)\, ,
\nonumber\\
\pi & = & g^{\mu\nu}\, \pi_{\mu\nu} = {\pi^\mu}_\mu = -\, \left( D - 2 \right)\, K
\, - \, \left( 2\, a_o\, D - 4\, a_o + b_o \right)\, \R\, K
\nonumber\\
& & \qquad\qquad\qquad\qquad
\, - \, \left( b_o\, D - 3\,  b_o + 4\, c_o \right)\, \R_{\mu\nu}\, K^{\mu\nu}\, 
+\, \O \left( K^3 \right)\,  ,
\label{zk}
\eeqa
where
\beqa
\dot{\phi^I} & = & \L_r\, \phi^I = {{d \phi^I}\over dr} \, ,
\nonumber\\
K_{\mu\nu} & = & {1\over 2}\, \L_r\, g_{\mu\nu} = -\, {1\over 2}\, \dot{g}_{\mu\nu} = 
-\, {1\over 2}\, {{d g_{\mu\nu}}\over{dr}}\, .
\label{zj}
\eeqa
Replacing $\dot{\phi^I}$ and $\dot{g}_{\mu\nu}$ in $\L_T$ by their canonical momenta,
and performing the Legendre transformation, we find the following expression for the 
total Hamiltonian
\beqa
-\,\, {1\over\sqrt{g}}\,\, \H_T & = & \left( \R + 2\, \Lambda + a_o\, \R^2 + b_o\, \R_{\mu\nu}\, \R^{\mu\nu} 
+\, c_o\, \R_{\mu\nu\rho\sigma}\, \R^{\mu\nu\rho\sigma} \right)
\nonumber\\
& & 
\, + \, \left( V \left(\phi\right)\, -\, 
{1\over 2}\, \nabla_\mu\, \phi^I\, \nabla^\mu\, \phi^I \right)
\nonumber\\
& & 
\, +\, \left( \pi_{\mu\nu}\, \pi^{\mu\nu} - {{\pi^2}\over{D - 2}} \right)
\, +\, {1\over 2}\, \pi_I\, \pi_I
\nonumber\\  
& &
\, -\, 2\, a_o\, \R\, \left( \pi_{\mu\nu}\, \pi^{\mu\nu} - {{\pi^2}\over{D - 2}} \right)
\, - \, 2\, b_o\, \R_{\mu\nu} \, \left( {\pi^{\mu}}_\rho\, \pi^{\rho\nu} - 
{{\pi\, \pi^{\mu\nu}}\over{D - 2}} \right)
\nonumber\\
& &
\, + \, 4\,\, c_o\, \R_{\mu\nu\rho\sigma}\, \pi^{\mu\rho}\, \pi^{\nu\sigma}
\, + \, {{4\, c_o}\over{\left(D - 2\right)^2}}\, \R\, \pi^2
\, -\, {{8\, c_o}\over{D - 2}}\, \R_{\mu\nu}\, \pi\, \pi^{\mu\nu}\, .
\label{zl}
\eeqa

\section*{Acknowledgments}

The author is grateful to Prof. W. G. Unruh for financial support, for encouragement and
for making the theory center a wonderful place where to work and make progress every day. 
I would like to thank him also for his valuable comments at various stages in the development
of this project. It is also a pleasure to acknowledge useful conversations with H. Verlinde and
B. Ovrut during the String Cosmology Workshop in July 2000. I would like also to thank the people of 
ICTP, in Trieste, for their hospitality and financial support at the early stage of this
project, in particular K. S. Narain and S. Randjbar-Daemi. Finally, I would like to thank all
the participants of the gravity weekly seminar and PIMS string journal club at UBC, in particular
E. T. Akhmedov, R. Scipioni, G. W. Semenoff, S. Surya and K. Zarembo.

\end{document}